\documentclass[pra,aps,showpacs,twocolumn,floatfix]{revtex4}
\usepackage{graphicx}
\usepackage{array}
\usepackage{color}
\usepackage{amsmath}
\usepackage{amsxtra}
\usepackage{amstext}
\usepackage{amssymb}
\usepackage{latexsym}
\usepackage{dsfont}
\usepackage{cases}
\usepackage{mathrsfs}
\usepackage{color}
\begin{document}
\title{Lie transformation on shortcut to adiabaticity in parametric driving quantum system}
\author{Jian-jian Cheng}
\author{Yao Du}
\author{Lin Zhang}\email{zhanglincn@snnu.edu.cn}
\affiliation{School of physics and information technology, Shaanxi Normal University,Xi'an 710119, P. R. China}
\begin{abstract}
{Shortcut to adiabaticity (STA) is a speed way to produce the same final state that
would result in an adiabatic, infinitely slow process. Two typical techniques to engineer
STA are developed by either introducing auxiliary counterdiabatic fields or finding new Hamiltonians
that own dynamical invariants to constraint the system into the adiabatic paths.
In this paper, a consistent method is introduced to naturally connect the above two
techniques with a unified Lie algebraic framework, which neatly removes the requirements
of finding instantaneous states in the transitionless driving method and the invariant
quantities in the invariant-based inverse engineering approach. The general STA schemes for different
potential expansions are concisely achieved with the aid of this method.}
\end{abstract}
\maketitle

\section{Introduction}
Quantum adiabatic technique is one of the most promising strategies for quantum computation based on
quantum adiabatic theorem \cite{Nielsen,Farhi}.
 ``Shortcuts to adiabaticity" (STA) are control protocols
that take the system quickly to the same populations, or even the same final states reached by slowly adiabatic
processes \cite{Mod}. A motivation to apply STA methods to quantum systems is to manipulate quantum states with a fast
coherent dynamics with high fidelity \cite{delCampo01}. Thus STA have become the typical techniques in preparing
or driving internal and motional states in atomic, molecular, optical, and solid-state physics
\cite{Mechanics, optical,transport,heating}.

Due to the time consuming problem of slowly parametric driving in adiabatic evolution, two equivalent strategies
for STA engineerings were developed so far, one is the transitionless quantum driving method \cite{tr1,tr2,tr3,tr4,tr5},
the other is the invariant-based inverse engineering technique\cite{x1,ph}.
The basic idea of transitionless quantum driving is to add an auxiliary interaction $\hat{H}_1(t)$ to reference Hamiltonian
$\hat{H}_{0}(t)$,
\begin{equation}
\hat{H}_{0}(t)=\sum_{n}|n(t)\varepsilon_{n}(t)\langle n(t)|,
\label{H0}
\end{equation}%
so that the dynamics follows exactly the approximate adiabatic evolution driven by $\hat{H}(t)$ starting
from $\hat{H}_{0}(0)=\sum_{n}|n(0)\rangle \varepsilon_n(0) \langle n(0)|$ through evolution operator
\begin{equation}
\hat{U}(t)=\sum_{n}|\psi_{n}(t)\rangle\langle n(0)|,
\end{equation}
where
\begin{equation}
\label{H01}
|\psi_{n}(t)\rangle=e^{-\frac{i}{\hbar}\int_{0}^{t}dt'\varepsilon_{n}(t')}|n(t)\rangle
\end{equation}
is the adiabatic states starting with $|n(0)\rangle$ without the geometric phase for the one parametric driving system.
The auxiliary interaction is then given by Berry's formulation \cite{X2}.
\begin{eqnarray}\label{h4t}
\begin{aligned}
\hat{H}(t)&&=& i\hbar(\partial_{t}\hat{U})U^{\dagger}  \\
&&=& \hat{H}_0(t)+ i\hbar\sum_{n}|\partial_{t}n(t)\rangle\langle n(t)|         \\
&& \equiv & \hat{H}_0(t)+\hat{H}_1(t)
\end{aligned}
\end{eqnarray}%
where $\hat{H}_{1}(t)$ is nondiagonal in the basis of ${|n(t)\rangle}$, and it requires a full knowledge
of spectral properties of the instantaneous Hamiltonian $\hat{H}_{0}(t)$. However, in most practical cases, the auxiliary interaction calculated by instantaneous
eigenstates are very hard to carry out.
Now we give an alternative method to do STA in a general parametric driving
system with Lie transformations.
Starting from a general harmonic oscillator Hamiltonian \cite{li1},
three independent generators can be separately defined by
\begin{equation}
\hat{J}_{+}=\frac{1}{2\hbar}\hat{x}^{2},\hat{J}_{-}=\frac{1}{2\hbar}\hat{p}%
^{2},\hat{J}_{0}=\frac{i}{4\hbar}\left( \hat{x}\hat{p}+\hat{p}\hat{x}%
\right),
\end{equation}%
where $\hat{J}_0$ is an anti-Hermitian operator for $\hat{J}^{\dagger}_0=-\hat{J}_0$. The three operators
satisfy the $\mathrm{SU}(2)$ commutation relations
\begin{equation}
[\hat{J}_{+},\hat{J}_{-}]=2\hat{J}_{0},[\hat{J}_{0},\hat{J}_{\pm}]=\pm\hat{J}_{\pm},
\end{equation}
and we can introduce three independent unitary transformations based on $\mathrm{SU}(2)$ by \cite{zhang}
\begin{equation}
\hat{U}_{\pm }(t)=e^{i\theta _{\pm }(t)\hat{J}_{\pm }},\hat{U}_{0}(t)=e^{2\theta
_{0}(t)\hat{J}_{0}},
\end{equation}%
where the transformation parameters $\theta _{\pm}(t)$ and $\theta _{0}(t)$
are piecewise continuous real functions defined within a control time
interval $[0,\tau]$. The above quantum transformations $\hat{U}_{\pm}(t)$ and $\hat{U}_{0}(t)$ correspond to classical canonical
transformations called rotations and squeezing, respectively. This type of unitary operators constitutes our
``Lie transformations", which provides a general method to construct the adiabatic Hamiltonian for STA designs.

\section{Parametrically driven quantum harmonic oscillator}
\label{sec02}

The first example is a fast harmonic trap expansion, which has been received
much attentions because of its fundamental and practical implications. It applies
to the cold atoms involving an adiabatic tuning of the system after a cooling phase \cite{cooling}.
For a parametric driven harmonic oscillator, we consider the system Hamiltonian
\begin{equation}
\hat{H}_{0}(t)=\frac{\hat{p}^{2}}{2m}+\frac{1}{2}m\omega ^{2}(t)\hat{x}^{2}=\hbar\omega(t)\left(\hat{a}_{t}^{\dag}\hat{a}_{t}+\frac{1}{2}\right),
\end{equation}%
where $\hat{a}_{t}$ and $\hat{a}^{\dag}_{t}$ are the annihilation and creation operators at time $t$ defined by
\begin{equation}%
\label{H1t}
\hat{a}_{t}=\sqrt{\frac{m\omega(t)}{2\hbar}}\left[\hat{x}+\frac{i}{m\omega(t)}\hat{p}\right],
\end{equation}
\begin{equation}%
\label{H1t}
\hat{a}_{t}^{\dagger}=\sqrt{\frac{m\omega(t)}{2\hbar}}\left[\hat{x}-\frac{i}{m\omega(t)}\hat{p}\right].
\end{equation}
Since the frequency depends on time, the instantaneous ladder operator $\hat{a}_{t}$ ($\hat{a}_{t}^{\dagger}$) will create (annihilate)
different instantaneous states at different times adapting to different frequencies.

Now, we consider STA control on the parametric oscillator from $t=0$ to $\tau$ by Lie transformations.
In this case, the squeezing operator can be written as
\begin{equation}%
\label{H1t}
\hat{U}_{0}(t)=e^{2(\ln r)\hat{J}_{0}}=e^{\frac{1}{2}(\ln r)[\hat{a}^{2}-(\hat{a}^{\dagger})^{2}]},
\end{equation}
where the subscript $t$ of $\hat{a}$ and $\hat{a}^{\dag}$ has been
dropped because the squeezing combination $\hat{a}^{2}-(\hat{a}^{\dag})^{2}$  is actually independent of time. Here, the introduced transformation parameter $r(t)=\sqrt{\frac{\omega(t)}{\omega(0)}}$ plays as a linear squeezing (adiabatic)
scaling factor, which gives (see Appendix\ref{app})
\begin{equation}
\hat{U}_{0}(t)|n(0)\rangle=|n(t)\rangle,
\label{rt}
\end{equation}%
and the instantaneous states of the final system satisfies
\begin{equation}%
\label{H1t}
\hat{H}_0(t)|n(t)\rangle =\varepsilon _{n}(t)|n(t)\rangle.
\end{equation}
A direct substitution of Eq.(\ref{rt}) into Eq.(\ref{h4t}) shows that the adiabatic Hamiltonian is
\begin{equation}
\begin{aligned}
\hat{H}(t) &&=&\sum_{n}\hat{U}_{0}|n(0)\rangle\varepsilon_{n}(t)\langle n(0)|\hat{U}_{0}^{\dagger}  \\
&&&+i\hbar\sum_{n}\dot{\hat{U}}_{0}|n(0)\rangle\langle n(0)|\hat{U}^{\dag}_{0} \\
&&=&r^{2}(t)\hat{U}_{0}\hat{H}_{0}(0)\hat{U}_{0}^{\dag}+i\hbar\dot{\hat{U}}%
_{0}\hat{U}^{\dag }_{0}  \\
&&=&\frac{\hat{p}^{2}}{2m}+\frac{1}{2}m\omega^{2} (t)\hat{x}^{2}-\frac{\dot{r}%
}{2r}(\hat{x}\hat{p}+\hat{p}\hat{x}),  \label{Hot}
\end{aligned}
\end{equation}
where the initial Hamiltonian $\hat{H}_0(0)=\frac{\hat{p}^{2}}{2m}+\frac{1}{2}m\omega^{2}_0\hat{x}^{2}$ with $\omega_0\equiv \omega(0)$.
Then the auxiliary Hamiltonian $\hat{H}_1(t)$ is given by
\begin{equation}%
\hat{H}_1(t)=-\frac{\dot{r}}{2r}(\hat{x}\hat{p}+\hat{p}\hat{x})%
=-\frac{\dot{\omega}(t)}{4\omega(t)}(\hat{x}\hat{p}+\hat{p}\hat{x}),
\label{H11w}
\end{equation}
which is firstly obtained by Berry's formula \cite{x2}.
This method provides a way to design the adiabatic driving of Eq.(\ref{h4t}) without resorting to
the instantaneous states of $\hat{H}_0(t)$. The result indicates that the higher speed of the state evolution,
the larger intensity of the required auxiliary field needs to
recover the adiabatic evolution.
The result clearly implies that, to an initial state described by a
linear superposition $\Psi(0)=\sum_{n}c_{n}|n(0)\rangle$, the evolution generated by adiabatic Hamiltonian $\hat{H}(t)$
leads to adiabatic state $\Psi(t)=\sum_{n}c_{n}|\psi_n(t)\rangle=\sum_{n}c_{n}|n(t)\rangle e^{-\frac{i}{\hbar }\int_{0}^{t}\varepsilon_n(t^{\prime}) dt^{\prime} }$.

However, the control field $\hat{H}_1(t)$ obtained by the above Lie transformation
is still unpractical and hard to be realized because it is non-local.
An alternative control field that is local and
experimentally realizable can be established by a successive transform of
\begin{equation}
\label{Up}
\hat{U}_{+}(t)=e^{i\theta_{+}(t) \hat{J}_{+}}, \quad \theta_{+}(t)=-\frac{m\dot{r}(t)}{r(t)},
\end{equation}
which continues to transform Hamiltonian $\hat{H}(t)$ into
\begin{equation}
\begin{aligned}
\hat{H}^{\prime }(t) &&=&\hat{U}_{+}\left( t\right) \hat{H}(t)\hat{U}%
_{+}^{\dag }\left( t\right) +i\hbar\dot{\hat{U}}_{+}\left( t\right) %
\hat{U}_{+}^{\dag }\left( t\right)    \\
&&=&\frac{\hat{p}^{2}}{2m}+\frac{1}{2}m\Omega ^{2}(t)\hat{x}^{2},  \label{HHt}
\end{aligned}
\end{equation}
where
\begin{equation}
\Omega^{2}(t)=\omega^{2}(t)-\frac{3}{4}\left[\frac{\dot{\omega}(t)}{\omega(t)}\right]^{2}
+\frac{1}{2}\frac{\ddot{\omega}(t)}{\omega(t)}
\end{equation}
is the modified time-dependent frequency that can be easily realized in a practical control\cite{x3}.

The new Hamiltonian $\hat{H}'(t)$ has the same structure as the reference
Hamiltonian $\hat{H}_{0}(t)$ but with different time-dependent frequency, which can be formed by magnetic and optical fields in the
cold atom experiments for a time-varying trapping.

Under the above successive transformations, the time evolution of
the initial state $|n(0)\rangle$ is mapped to
\begin{equation}
\label{phit}
|\Phi_{n}(t)\rangle=\hat{U}_{+}(t)|\psi_{n}(t)\rangle=e^{-i\frac{m\dot{\omega}(t)}{4\hbar\omega(t)}x^{2}}|\psi_{n}(t)\rangle,
\end{equation}
which generates a similar final state of an adiabatic process
evolving slowly from $\hat{H}_{0}(0)$ to $\hat{H}_{0}(\tau )$ if the control frequency $\Omega(t)$
meets the boundary conditions%
\begin{equation}
\Omega(0)=\omega(0)\equiv\omega_{0},\quad \Omega(\tau )=\omega(\tau)\equiv\omega_{\tau}. \label{Ome}
\end{equation}%
The above conditions lead to
\begin{equation}
\dot{\omega}(0)=\dot{\omega}(\tau )=0,\quad \ddot{\omega}(0)=\ddot{\omega}(\tau
)=0.  \label{omega}
\end{equation}
The zero of the first time-derivative means that $|\Phi_{n}(0,\tau)\rangle=|\psi_{n}(0,\tau)\rangle$ and the zero of the second one implies
$\hat{H}'(0,\tau)=\hat{H}_0(0,\tau)$.

In order to design the frequency $\Omega(t)$ to meet Eq.(\ref{omega}), a polynomial ansatz of time function
for $\omega(t)$ is often used \cite{x1}
\begin{equation}
\omega(t)=\omega_{0}+10\delta s^{3}-15\delta s^{4}+6\delta s^{5},
\end{equation}
where $\delta=\omega_{\tau}-\omega_{0}$ and $s=t/\tau$.
\begin{figure}[htbp]
\begin{center}
\includegraphics[width=220pt]{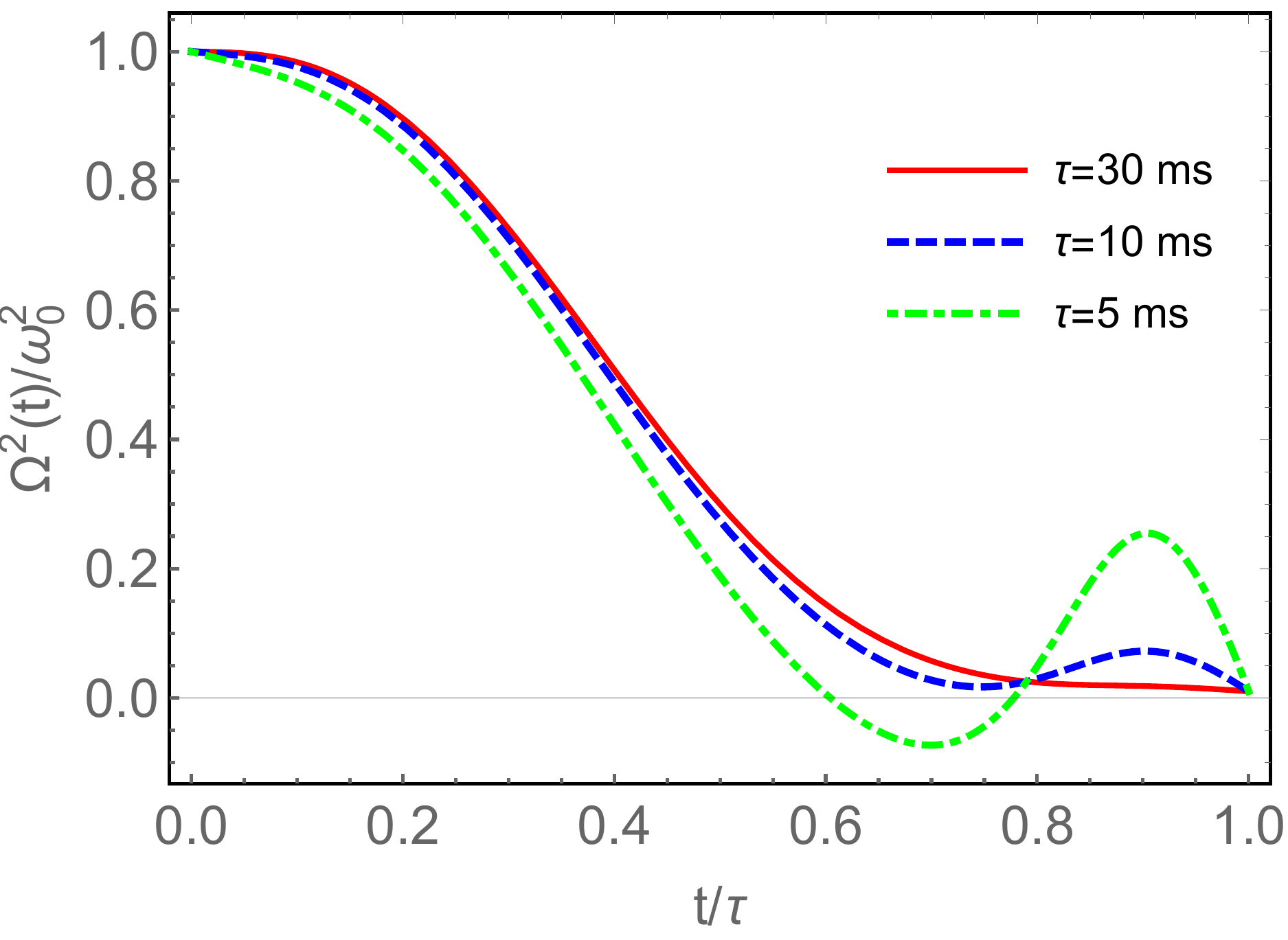}
\end{center}
\caption{The adiabatic drivings of the trap frequency $\Omega^{2}(t)$ changing
from $\omega_{0}=2\pi\times 250$ Hz to $\omega_{\tau}=2\pi\times25$ Hz
are displayed for three different control times
$\tau=30$ ms (red solid line), $\tau=10$ ms (blue dashed line) and
$\tau=5$ ms (green dot-dashed line).}
\label{fig1}
\end{figure}
The trajectories with different control times are displayed in Fig.\ref{fig1}.
We can see $ \Omega^{2}(t)$ may become negative during some time intervals if the control process
is very fast (green dot-dashed line), making the potential becomes an expulsive parabola. These state-independent shortcuts are ideal for
maximal robustness owing to no excitation in the final state but allowing
for excitations in the intermediate.
With Lie transformation, it is convenient to verify that the above STA design is equivalent to the Lewis-Riesenfeld (LR) invariants method\cite{x1}.
If we set the total transformation $\hat{U}_{I}(t)=\hat{U}_{+}(t)\hat{U}_{0}(t)$,
then the new adiabatic Hamiltonian $\hat{H}^{\prime}(t)$ is obtained by
\begin{equation}%
\label{Hi11}
\hat{H}^{\prime}(t)=r^{2}\hat{I}(t)+i\hbar\dot{\hat{U}}_{I}(t)\hat{U}_{I}^{\dag}(t),
\end{equation}
where the LR invariant is defined by
\begin{equation}
\label{H27}
\hat{I}(t)=\hat{U}_{I}(t)\hat{H}_{0}(0)\hat{U}_{I}^{\dag}(t).
\end{equation}
Naturally, we can see $\hat{I}(0)=\hat{H}_{0}(0)$
and it satisfies
\begin{equation}%
\frac{d \hat{I}(t)}{dt}=\frac{\partial \hat{I}(t)}{\partial t}-\frac{1}{i\hbar}[\hat{H}^{\prime}(t),\hat{I}(t)]=0.
\end{equation}
From Eq.(\ref{H27}), we can find the eigenvectors of $\hat{I}(t)$ are
\begin{equation}%
\label{Ieign}
|\phi_{n}(t)\rangle=\hat{U}_{I}(t)|n(0)\rangle,
\end{equation}
which satisfy
\begin{equation}%
\label{H1t}
\hat{I}(t)|\phi_{n}(t)\rangle=\lambda_{n}|\phi_{n}(t)\rangle,
\end{equation}
where $\lambda_{n}=(n+\frac{1}{2})\hbar\omega_{0}=\varepsilon_{n}(0)$.
According to the invariant-based inverse engineering, the orthonormal eigenvectors of the
invariant $\hat{I}(t)$ and the instantaneous eigenstates of system $\hat{H}_0(t)$ coincide at the initial and final times, then we naturally have
\begin{eqnarray}
\hat{I}(t)=\sum_{n}|\phi_{n}(t)\rangle\lambda_{n}\langle \phi_{n}(t)|
=\sum_{n}|n(t)\rangle\lambda_{n}\langle n(t)|,
\end{eqnarray}
which leads to
\begin{equation}%
\label{H1t}
\hat{I}(t)=r^{-2}\hat{H}_{0}(t)=\frac{\omega_{0}}{\omega(t)}
\left[\frac{\hat{p}^{2}}{2m}+\frac{1}{2}m\omega^{2}(t)\hat{x}^{2}\right].
\end{equation}
As the new Hamiltonian of Eq.(\ref{Hi11}) must be equivalent to the
Hamiltonian of the system $\hat{H}_{0}(t)$ at the initial and
final times for STA controls, the boundary conditions will be
\begin{equation}%
\dot{\hat{U}}_{0}(t={0,\tau})=0 ,\dot{\hat{U}}_{+}(t={0,\tau})=0.
\label{Hinii}
\end{equation}
This new constraints can naturally recover Eq.(\ref{omega}) and make $\hat{H}_{1}(t)$ vanish at the initial and final times.
It should be noted that the transformed Hamiltonian $\hat{H}'(t)$ we found here is just the Hamiltonian for LR dynamical invariant $\hat{I}(t)$. This is the essence of
invariant-based inverse engineering and our method avoids the difficulty for constructing dynamical
invariants $\hat{I}(t)$ for a given system \cite{li3}.

In conclusion, complementary to existing approaches about harmonic oscillator expansions, the Lie algebraic method proposed here provides a more general unified framework for STA contorls.\\
\section{Power-law trap}

In the Lie algebraic framework, a generation to other trapping potentials is convenient and direct.
We begin our trick with a free particle inside an expanding one-dimensional box  (quantum piston) \cite{box1}
\begin{equation}%
\hat{H}_{0}(t)=\frac{\hat{p}^{2}}{2m}+V_{\mathrm{box}}(x,t),
\end{equation}
where the potential is
\begin{equation}
\label{box}
V_{\mathrm{box}}(x,t)=
\left\{
\begin{array}{rcl}
0, &0<x<L(t)\\
\infty,&\mathrm{otherwise}
\end{array} \right.
\end{equation}
which describes an infinite square well with a moving boundary at $x=L(t)$.
If the expansion is slow enough to follow adiabatic theorem, the exact solution can
be expressed as
\begin{equation}
\begin{aligned}
\psi_{n}(x, t) &&=&\sqrt{\frac{2}{L(t)}}\sin\left[\frac{n\pi %
x}{L(t)}\right]e^{-\frac{i}{\hbar}\int_{0}^{t}dt'\frac{n^{2}\pi^{2}\hbar^{2}}{2mL^{2}(t)}}    \\
&&\equiv&\varphi_{n}\left[\frac{x}{L(t)}\right]e^{-\frac{i}{\hbar }\int_{0}^{t}\varepsilon_n(t^{\prime}) dt^{\prime} },  \label{HHt}
\end{aligned}
\end{equation}
where the instantaneous states satisfy
\begin{equation}
\hat{H}_0(t)\varphi_n\left[\frac{x}{L(t)}\right]=\varepsilon_n(t)\varphi_n\left[\frac{x}{L(t)}\right].
\end{equation}

Since the width of box depends on time, a squeezing transformation can be introduced
\begin{equation}
\hat{U}_{0}(t)=e^{2\ln\frac{L(0)}{L(t)}\hat{J}_{0}},
\end{equation}
with an initial width $L(0)\equiv L_{0}$. In the coordinate representation, the squeezing means increasing the wavelength of the free particle and reducing its momentum,
which gives
\begin{equation}%
\label{Hinit}
\hat{U}_{0}(t)\varphi_{n}\left(\frac{x}{L_{0}}\right)=\varphi_{n}\left[\frac{x}{L(t)}\right].
\end{equation}
By using the same STA design method, we have
\begin{equation}
\label{Hwx}
\begin{aligned}
\hat{H}(t) &&=&\sum_{n}\hat{U}_{0}|\varphi_{n}(\frac{x}{L_{0}})\rangle\varepsilon_{n}(t)\langle \varphi_{n}(\frac{x}{L_{0}})|\hat{U}_{0}^{\dagger} \\
&&+&i\hbar\sum_{n}\dot{\hat{U}}_{0}|\varphi_{n}(\frac{x}{L_{0}})\rangle\langle \varphi_{n}(\frac{x}{L_{0}})|\hat{U}_{0}^{\dagger} \\
&&=&\frac{\hat{p}^{2}}{2m}+\frac{\dot{L}(t)}{2L(t)}(\hat{x}\hat{p}+\hat{p}\hat{x}).
\end{aligned}
\end{equation}
The above Hamiltonian can be used to control quantum gas of noninteracting particles confined in an expanding box with the initial
populations sampled microcanonically. The gas will remain a uniform distribution for an arbitrary expanding boundary $L(t)$
because the auxiliary term in Eq.(\ref{Hwx}) suppresses shock waves (excitations) by uniformly expanding the gas \cite{box2}.

This technique can be easily generalized to all even-power-law potentials \cite{box3}
\begin{equation}%
\label{Hinit}
\hat{V}(r x_{1}, r x_{2},\cdots,r x_{n})=r^{k}\hat{V}(x_{1},x_{2},\cdots,x_{n}),
\end{equation}
with $k=2,4,6\cdots$ due to the trapping properties of the potentials. Clearly, the case $k=2$ is
the harmonic trapping we have considered in section II and $k\rightarrow\infty$ corresponds to the quantum box.
The virial theorem shows that the average kinetic energy of the particle in the above potential
$\overline{T}_n=\frac{k}{k+2}\bar{\varepsilon}_n\equiv\mu\bar{\varepsilon}_n$ \cite{ld}.

For convenience, we set $L_{0}=1$ and the squeezing operator becomes $\hat{U}_{0}(t)=e^{-\mu\ln [L(t)]\hat{J}_{0}}$, which leads to
\begin{equation}%
\hat{U}_{0}(t)\varphi_{n}(x)=\varphi_{n}\left[\frac{x}{L^{\mu}(t)}\right], \label{U0t}
\end{equation}
where $\varphi_{n}(x)$ are the wavefunctions of Hamiltonian with a one-dimensional potential of Eq.(\ref{Hinit}).
Substitute Eq.(\ref{U0t}) into Eq.(\ref{h4t}), we can find the controlling adiabatic Hamiltonian
\begin{equation}%
\hat{H}(t)=\hat{H}_{0}(t)+\frac{k}{k+2}\frac{\dot{L}(t)}{2L(t)}(\hat{x}\hat{p}+\hat{p}\hat{x}).\label{HHHt}
\end{equation}
It shows that squeezing transformation greatly facilitates the STA designs for expansion or compression of the homogeneous trapping potentials.
But still the auxiliary term is difficult to implement
in the laboratory, we conduct another transformation $\hat{U}_{+}(t)=\exp\left[\frac{im\dot{\xi}}{2\hbar\xi}\hat{x}^{2}\right]$ with $\xi=L^{\mu}(t)$ on Eq.(\ref{HHHt}) and arrive
\begin{equation}%
\hat{H}^{\prime}(t)=\hat{H}_{0}(t)-\frac{1}{2}m\frac{k}{k+2}\frac{\ddot{L}(t)}{L(t)}x^{2}.
\end{equation}
The above new Hamiltonian introduces an auxiliary potential
\begin{equation}%
U_{\mathrm{aux}}(x,t)=-\frac{1}{2}m\frac{k}{k+2}\frac{\ddot{L}(t)}{L(t)}x^{2}
\end{equation}
to control the system to a final adiabatic state.
Now the auxiliary potential is a realistic harmonic potential which can be implemented in the laboratory with well-established
technology and have been applied to perform a fast adiabatic squeezing in Bose-Einstein condensate \cite{Bose}.

\begin{figure}[htpb]
\begin{center}
\includegraphics[width=220pt]{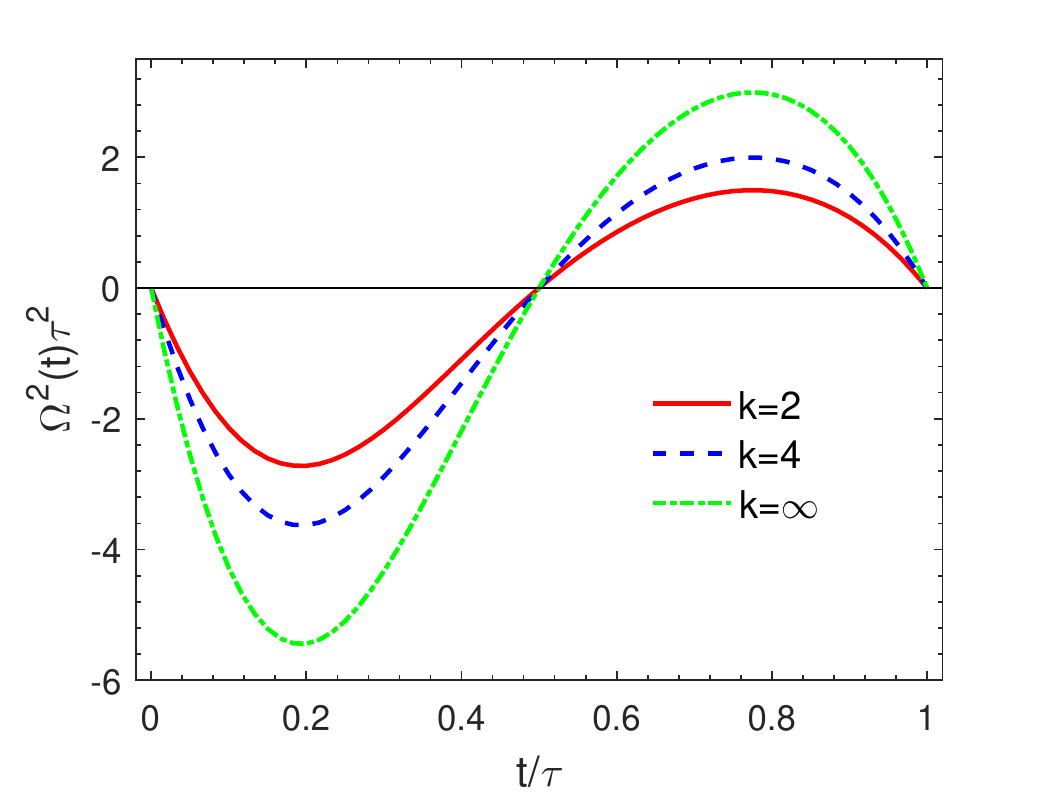}
\end{center}
\caption{Shortcuts to adiabaticity by adiabatic driving, for an
expansion factor $L(\tau)=2$ with different power-law traps. The exponent values of auxiliary  potentials correspond to harmonic oscillator (red solid line), a quartic trap potential (blue dashed line), and an expanding quantum piston (green dot-dashed line).}
\label{fig2}
\end{figure}
In this case, the LR invariant in the inverse-engineering approach will also be defined by $\hat{I}(t)=\hat{U}_{I}(t)\hat{H}_{0}(0)\hat{U}_{I}^{\dagger}(t)$. Here, the effective frequency of the auxiliary potential, $\Omega^{2}(t)=-\frac{k}{k+2}\frac{\ddot{L}(t)}{L(t)}$, is the key to engineer the STA. According to Eq.(\ref{Hinii}), we impose the boundary conditions $\dot{L}(0,\tau)=\ddot{L}(0,\tau)=0$. A polynomial ansatz to meet the boundary condition can be designed by
\begin{equation}%
L(t)=1+[L(\tau)-1][10 s^{3}+3s^4(2s-5)].
\end{equation}%
The driving frequencies $\Omega^2(t)$ for different power-law potentials are displayed in Fig.\ref{fig2}. It shows a symmetric control process and $U_{\mathrm{aux}}(x,t)$ performs a repulsive harmonic potential in an early stage ($ t<\frac{\tau}{2}$), achieving a speed-up STA in an arbitrary
finite time $\tau$. In the subsequent stage $(t>\frac{\tau}{2})$, $\Omega^{2}(t)$ changes its sign to positive and $U_{\mathrm{aux}}(x,t)$ becomes a trapping potential, slowing down the expanding mode. In other words, when the intermediate state deviates from the target state, the auxiliary potential gives a negative feedback, which reproduces the adiabatic result at
the end of the evolution. Fig.\ref{fig2} also reveals that a stronger trapping potential means a higher auxiliary control energy.
This conclusion has been considered in the expansion stroke of a quantum piston \cite{box4}, where an auxiliary harmonic trap is introduced by an electromotive circuit which
dissipates a mass of energy to preform the speed-up adiabatic Otto circle.

\section{Conclusion}
A unified algebraic framework is developed to design
shortcuts to adiabaticity in both transitionless quantum driving and invariant-based inverse engineering method.
We demonstrate that the two approaches is mathematically equivalent and share a common physical ground with Lie transformations, and provides a universal method to design
STA for the general time-dependent parametric driving systems.
It removes the demanding requirement to diagonalize the Hamiltonian of the system in the transitionless driving method, and the difficulty to find the invariant quantities in the invariant-based inverse engineering technique.
The Lie transformation method present a powerful tool to design STA control in the homogenous power-law trapping system and can neatly connect the invariant-based
method with the transitionless driving method by quickly diagonalizing the reference Hamiltonian.
In summary, this work provides a deeper insight of shortcut-to-adiabaticity techniques in the quantum control problems.

\section*{ACKNOWLEDGMENTS}
This work was supported by the National Natural Science Foundation of China for
emergency management project (Grant No.11447025, 11847308).

\section*{APPENDIX: THE DERIVATIONS OF SQUEEZING TRANSFORMATION}
\label{app}
The squeezing transformation is a unitary operator and easy to prove that
~\\
~\\
~\\
\begin{equation}
\renewcommand\theequation{A.1}
\begin{aligned}
\hat{U}_{0}(t)\hat{a}_{0}\hat{U}^{\dagger}_{0}(t)&&=&\hat{a}_{0}\cosh(-\ln r)-\hat{a}^{\dag}_{0}\sinh(-\ln r)  \\
&&=&\sqrt{\frac{m\omega_{0}}{2\hbar}}\left[\hat{x}+\frac{i}{m\omega_{0}}\hat{p}\right]\frac{1+r^{2}}{2r} \\
&&-&\sqrt{\frac{m\omega_{0}}{2\hbar}}\left[\hat{x}-\frac{i}{m\omega_{0}}\hat{p}\right]\frac{1-r^{2}}{2r}\\
&&=&\sqrt{\frac{m\omega(t)}{2\hbar}}\left[\hat{x}+\frac{i}{m\omega(t)}\hat{p}\right]=\hat{a}_{t},
\end{aligned}
\end{equation}
\begin{equation}
\renewcommand\theequation{A.2}
\begin{aligned}
\hat{U}_{0}(t)\hat{a}^{\dagger}_{0}\hat{U}^{\dag}_{0}(t)&&=&\hat{a}^{\dag}_{0}\cosh(-\ln r)-\hat{a}_{0}\sinh(-\ln r)  \\
&&=&\sqrt{\frac{m\omega_{0}}{2\hbar}}\left[\hat{x}-\frac{i}{m\omega_{0}}\hat{p}\right]\frac{1+r^{2}}{2r} \\
&&-&\sqrt{\frac{m\omega_{0}}{2\hbar}}\left[\hat{x}+\frac{i}{m\omega_{0}}\hat{p}\right]\frac{1-r^{2}}{2r} \\
&&=&\sqrt{\frac{m\omega(t)}{2\hbar}}\left[\hat{x}-\frac{i}{m\omega(t)}\hat{p}\right]=\hat{a}^{\dag}_{t},
\end{aligned}
\end{equation}
where $\hat{a}_{0}\equiv \hat{a}(0)$ and $\hat{a}_{t}\equiv \hat{a}(t)$.
So that
\begin{equation}
\renewcommand\theequation{A.3}
\begin{aligned}
\hat{U}_0(t)|n(0)\rangle &&=&\hat{U}_0(t)\frac{1}{\sqrt{n!}}(\hat{a}_{0}^{\dagger})^{n}|0_{0}\rangle   \\
&&=&\frac{1}{\sqrt{n!}}\hat{U}_0(t)(\hat{a}_{0}^{\dagger})^{n}\hat{U}^{\dagger}_0(t)\hat{U}_0(t)|0_{0}\rangle  \\
&&=&\frac{1}{\sqrt{n!}}\left[\hat{U}_0(t)\hat{a}_{0}^{\dagger}\hat{U}^{\dagger}_0(t)\right]^{n}\hat{U}_0(t)|0_{0}\rangle  \\
&&\equiv &\frac{1}{\sqrt{n!}}(\hat{a}_{t}^{\dagger})^{n}|0_{t}\rangle=|n(t)\rangle,
\end{aligned}
\end{equation}
where the vacuum state at time $t$ is defined by $|0_{t}\rangle\equiv\hat{U}_{0}(t)|0_{0}\rangle$.

\end{document}